\documentclass[12pt]{article}
\begin{document}
\thispagestyle{empty}
\noindent\hspace*{\fill}  FAU-TP3-00/4 \\
\noindent\hspace*{\fill}  hep-th/0007081 \\
\noindent\hspace*{\fill}  \today   \\

\begin{center}\begin{Large}\begin{bf}
Chiral Condensates in the  Light-Cone Vacuum \\
\end{bf}\end{Large}\vspace{.75cm}
 \vspace{0.5cm}
Frieder Lenz and Michael Thies,\\
\vskip 0.2cm
{\em Institut f\"ur Theoretische Physik III \\
Universit\"at Erlangen-N\"urnberg \\
Staudtstra{\ss}e 7 \\
D-91058 Erlangen, Germany} \\
\vskip 0.4cm
Koichi Yazaki,\\
\vskip 0.2cm
{\em College of Arts and Sciences,\\
Tokyo Woman's Christian University,\\
2-6-1 Zenpukuji,\\
Suginami-ku, Tokyo 167-8585, Japan}
\end{center}
\vspace{1cm}\baselineskip=35pt

\date{\today}
\begin{abstract} \noindent
In light-cone quantization, the standard procedure to characterize the phases of a system
by appropriate ground state expectation values fails. The light-cone vacuum is determined
kinematically. We show that meaningful quantities which can serve as order parameters
are obtained as expectation values of Heisenberg operators in the equal (light-cone)
time limit. These quantities differ from the purely kinematical expectation values of the
corresponding Schr\"odinger operators. For the Nambu--Jona-Lasinio and the
Gross-Neveu model, we describe the spontaneous breakdown of chiral symmetry;
we derive within light-cone quantization the corresponding gap equations and the values
of the chiral condensate.
\end{abstract}

\newpage\baselineskip=18pt

Inherent to the light-cone description of quantum field theories is the triviality of the vacuum.
Most of the simplifying features of  light-cone quantization as well as foundation and phenomenological
success of the quark-parton model are, to a large extent, related to the simplicity of the structure of the
vacuum (cf. the reviews on light-cone quantization \cite{Burkardt96,BPP98}). The simplicity of
the vacuum is independent of dynamics, it is of kinematical origin. In the light-cone formulation,
Minkowski space-time is described by the metric
\begin{equation}
  \label{lc1}
g_{\mu\nu}=\left (\begin {array}{rrrr} 0&1&0&0\\\noalign{\medskip}1&0
&0&0\\\noalign{\medskip}0&0&-1&0\\\noalign{\medskip}0&0&0&-1
\end {array}\right )\end{equation}
and  parametrized by the coordinates
$$x^{\pm}=\frac{1}{\sqrt{2}}(x^{0} \pm x^{3}) \ , \qquad
x^{\perp}=(x^{1},x^{2}) \ .$$
With the form (\ref{lc1}) of the metric, the dispersion relation $p^{2}=m^{2}$ leads to the
following relation between the {\em light-cone energy} $p_{+}$ and momentum
components $p_{-},p_{\perp}$
\begin{equation}
  \label{lc3}
  p_{+} = \frac{p_{\perp}^{2}+m^{2}}{2p_{-}} \ .
\end{equation}
In contradistinction to  the standard parametrization of space-time, the light-cone energy $p_{+}$
assigned to a single particle state of a given momentum is unique. The sign of the energy is determined
by the sign of the momentum component $p_{-}$. Thus  in the absence of interactions, the fermionic
vacuum consists of occupied states with negative $p_{-}$  and of empty states with positive $p_{-}$.
This vacuum structure does not change when turning on interactions between the fermions.
No other states with equal momentum are available which could be reached by collisions
among the fermions. Thus the structure of the vacuum is independent of interactions.\\
This triviality of the vacuum  poses conceptual problems when applying light-cone quantization to
systems which are known to possess a non-trivial vacuum structure induced, for instance, by
spontaneous symmetry breakdown, Higgs mechanism or topological properties.
While the equivalence of light-cone quantization with more standard quantization has been
established perturbatively (cf. \cite{BPP98}) the triviality problem points to a lack of
understanding of this quantization scheme in the non-perturbative regime. It remains to be
understood how, in light-cone quantization,  different phases of a system can be built on a
vacuum which is determined kinematically. In particular, vacuum expectation values (VEV)
such as the chiral condensate $\langle 0 |\bar{\psi}\psi|0\rangle$ are trivial in light-cone
quantization and thus cannot serve as order parameters characterizing the realization of
symmetries. On the other hand, it is known from the study of low dimensional systems
such as the 't~Hooft model \cite{tHooft74} that light-cone quantization can reproduce correctly
spectra which
contain Goldstone bosons; furthermore, by using properties of the spectrum, the
correct value of the quark condensate could be determined \cite{Zhitnitsky} although
explicit calculation yields a vanishing VEV. \\
To clarify the physical relevance of the light-cone vacuum we
consider model theories in which spontaneous symmetry breakdown
of a continuous symmetry  occurs with the ensuing emergence of
Goldstone particles and formation of condensates. In the
Nambu--Jona-Lasinio model (NJL) \cite{NJL61} and its two dimensional
version, the (chiral) Gross-Neveu model (GN) \cite{GN74}, the breakdown of the chiral symmetry
is induced by mass generation of the fermions. The Lagrangian of these models has the following
structure
$${\cal L}= \bar{\psi}({\rm i}\partial_{\mu}\gamma^{\mu}-m)\psi+{\cal L}_{\rm int}
(\psi,\bar{\psi}) \ .$$
${\cal L}_{\rm int}$ is a 4-fermion self interaction.
This expression contains implicitly a sum over  fermion species (``color'') while flavor
dependences important in phenomenological applications are of no importance for our discussion.
In the following we shall display the formalism for the 3+1 dimensional NJL model and we shall
discuss later the necessary modifications for the lower-dimensional GN model.
We use a representation of the $\gamma$ matrices in which $\gamma_{5}$ and the projection
operators $\Lambda^{\pm}$ are given by
\begin{equation}
\label{gamma}
  \gamma_{{5}}=\left (\begin {array}{cc} \sigma_{3}&0\\\noalign{\medskip} 0&\sigma_{3}
\end {array}\right )\ ,\quad \Lambda^{\pm}=\frac{1}{2}(1\pm \gamma^{0}\gamma^{3})\ ,\quad
 \gamma^{0}\gamma^{3}=\left (\begin {array}{rr} {\bf 1}&0\\\noalign{\medskip} 0&{\bf -1}
\end{array}\right )\ .
\end{equation}

The projection operators  $\Lambda^{\pm}$ decompose the 4-spinor into 2-spinors
\begin{displaymath}
\psi = 2^{-1/4}\left (\begin {array}{c} \varphi\\ \noalign{\medskip}\chi
\end {array}\right )
\end{displaymath}
and the Lagrangian becomes
\begin{equation}
  \label{lc4}
{\cal L} =
{\rm i}\varphi^{\dagger}\partial_{+}\varphi+{\rm i}\chi^{\dagger}\partial_{-}\chi+
\frac{{\rm i}}{\sqrt{2}}\left(\varphi^{\dagger}\tilde{\partial}_{m}\chi+\chi^{\dagger}
\partial_{m}\varphi\right)+{\cal L}_{\rm int}(\varphi,\chi)
\end{equation}
with
$$
{\rm i}\partial_{m}={\rm i}\sigma_{3}\partial_{1}-\partial_{2}+\sigma_{1}m\ , \quad {\rm i}
\tilde{\partial}_{m}={\rm i}\sigma_{3}\partial_{1}+\partial_{2}+\sigma_{1}m\ .$$
Only the spinor $\varphi$ is dynamical, no time derivative of $\chi$ is present. In canonical
quantization, $\chi$ is treated as a constrained field. This reduction in the number of dynamical
degrees of freedom makes the single particle states with given momentum unique and thereby the
light-cone vacuum trivial.
In the representation  (\ref{gamma}), chiral rotations are defined by
\begin{equation}
  \label{chirot}
  \varphi(x)\rightarrow {\rm e}^{{\rm i}\alpha \sigma_{3}}\varphi(x)\ ,\quad \chi(x)\rightarrow
{\rm e}^{{\rm i}\alpha \sigma_{3}}\chi(x)\ .
\end{equation}
With the following choice of the 4-fermion interaction,
\begin{equation}
 \label{ NJL}
 {\cal L}_{\rm int} = \frac{g^2}{2}\left((\bar{\psi}\psi)^{2} +({\rm i}\bar{\psi}\gamma_{5}
\psi)^2\right)
=\frac{g^2}{4}\left(\left(\varphi^{\dagger}\sigma_{1}\chi
  +\chi^{\dagger}\sigma_{1}\varphi\right)^2+\left(\varphi^{\dagger}\sigma_{2}\chi
  +\chi^{\dagger}\sigma_{2}\varphi\right)^2\right)\ ,
\end{equation}
the NJL-Lagrangian is invariant under chiral rotations provided  the (bare) mass $m$ vanishes.
At this point we do not follow the standard path in employing  the canonical formalism; the
description in terms of  light-cone  Schr\"odinger operators will turn out to be too restrictive.
We rather study this model by using functional techniques based on the generating functional
\begin{equation}
  \label{gefu}
  Z [ \eta,\gamma] = \int D [\varphi, \chi ]  {\rm e}^{{\rm  i} \int {\rm d}^4x\,({\cal L}+
\varphi^{\dagger} \eta + \eta^{\dagger} \varphi + \chi ^{\dagger} \gamma + \gamma^{\dagger} \chi ) }\ .
\end{equation}
Since fermionic mass generation is the mechanism which drives the system into the spontaneously
broken phase the correlation function related to the chiral condensate for the case of
noninteracting ($g=0$) massive fermions reveals the difficulties in describing non-trivial vacua.
We consider
\begin{eqnarray}
  \label{ corr}
  C(x)&=&\langle 0|T(\varphi^{\dagger}(x)\sigma_{1}\chi(0))| 0 \rangle= {\rm i}m\,
2^{3/2}\int \frac{{\rm d}^{4}p}{(2 \pi)^{4}}\frac{{\rm e}^{{\rm i}px}}{ p^{2} -m^{2}
+{\rm i} \epsilon}\label{C1}\\
&=&  m\sqrt{2}  \left ( \frac{1}{2\pi}
    \right )^{3} \int {\rm d}^{2} p_{\perp} \int^{\infty}_{0} \frac{{\rm d}p_{-}}{p_{-}}
{\rm e}^{- {\rm i} \frac{p^{2}_{\perp} + m^{2} - {\rm i} \epsilon}{2 p_{-}} |x^{+}|+
 {\rm i} p_{\perp}x^{\perp}-{\rm i} p_{-}x^{-} \epsilon (x^{+})}\label{C2}\\
& = &\frac{1}{\sqrt{2} \pi ^{2}} \; \frac{m^{2}}
     {\sqrt{- x^{2} }} \; K_{1} (m\sqrt{-x^{2}})\label{C3} \ .
\end{eqnarray}
As has been noted quite some time ago \cite{Nakanishi77} in a discussion of bosonic theories,
values of such correlation functions are actually not well defined. In particular  evaluating
$C(x)$ for $x^{+}=0,$  using  Eq. (\ref{C2}) yields
\begin{equation}
  \label{sch1}
C_{\rm S}(x^{-},x^{\perp})=  m\sqrt{2}  \left ( \frac{1}{2\pi}
    \right )^{3} \int {\rm d}^{2} p_{\perp} \int^{\infty}_{0} \frac{{\rm d}p_{-}}{p_{-}}
{\rm  e}^{{\rm i} p_{\perp}x^{\perp} - {\rm i} p_{-}x^{-} }
\end{equation}
while using Eq. (\ref{C3})
\begin{equation}
  \label{hei1}
C_{\rm H}(x^{-},x^{\perp}) =\frac{1}{\sqrt{2} \pi ^{2}} \; \frac{m^{2}}
     {\sqrt{ x_{\perp}^{2} }} \; K_{1} (m\sqrt{x_{\perp}
^{2}})\ .
\end{equation}
Expression (\ref{sch1}) agrees with the result of the canonical formalism in which Schr\"odinger
operators are used.  This expression has only a trivial dependence on $m$, reflecting the triviality
of the vacuum. It is divergent even off the light-cone. On the other hand, the expression (\ref{hei1})
is regular for space- or timelike separations and depends non-trivially on the fermion mass. Furthermore
it is invariant under Lorentz transformations. The origin of this different behavior is a direct
consequence of the
light-cone dispersion relation. However small $x^{+}$ is chosen, there are always
states with sufficiently small $p_{-}$ available which give rise to oscillations in the
integrand in  (\ref{C2}) and thereby regularize the $1/p_{-}$ singularity. In standard
coordinates such an effect does not exist, $x^{0}=0$ can be chosen at every level
of the calculation and the result agrees with Eq. (\ref{hei1}).
From these observations we conclude: Expectation values of Schr\"odinger operators in the
light-cone vacuum do not agree with the limit of  expectation values of Heisenberg operators
\begin{equation}
\label{schei}
\lim_{x^{+}\rightarrow 0}\, \langle 0|T (\varphi ^{\dagger}(x)\sigma_{1} \chi(0) ) | 0 \rangle \neq
\langle 0 | \varphi ^{\dagger} (x^{+} = 0^{+}, x^{-}, x^{\perp} )\sigma_{1} \chi(0)  | 0 \rangle \ .
\end{equation}
Although we have computed these expectation values for non-interacting fermions, it is easy to
see that these arguments are essentially not changed when interactions are present. The triviality
of the vacuum implies that VEV's of Schr\"odinger operators do not change when including
interactions; on the other hand the absence of singularities in $C(x)$ for arbitrary small but
non-vanishing $ x^{+}$ and $ x^{2} \neq 0$ is easily demonstrated by inserting a complete
set of states (subtleties may only occur in 1+1 dimensional systems,
if massless particles are present.)
Furthermore, covariance dictates that in the absence of singularities,  vacuum expectation values
of Heisenberg operators at given spacelike $x^{2}$ are the same for $x^{+}\rightarrow 0$ and
$x^{0}= 0$
\begin{displaymath}
\lim_{x^{+}\rightarrow  0}\langle 0 | T (\varphi ^{\dagger}(x)\sigma_{1} \chi(0) ) | 0\rangle
\Big|_{x^{2}}=\langle  0 | \varphi ^{\dagger} (x^{0} = 0^{+},{\bf x} )\sigma_{1} \chi(0)  | 0
\rangle\Big|_{{\bf x}^{2}=-x^{2}}
\end{displaymath}
and coincide with the VEV of the $x^{0}= 0$ Schr\"odinger operators. Thus, on the light-cone,
VEV's of Heisenberg operators in the equal light-cone limit and not VEV's of Schr\"odinger
operators are physically meaningful quantities; in particular they can serve in the limit $x^{2}
\rightarrow 0$ as order parameters to characterize the phases of a system and properly define
for finite $x^{2}$ ``observable'' correlation functions.\\
We now demonstrate in a schematic light-cone calculation for the NJL model the
procedure for computing  condensate values. In the first step, the spectrum  of the light-cone
Hamiltonian has to be determined. In the above model this step is done easily for large $N$.
Replacing  in this limit the bilinear $(\chi^{\dagger}\sigma_{1}\varphi)$ by a $c$-number
\begin{equation}
 \label{mass}
 g^{2}\sum_{i=1}^{N}\chi^{\dagger}_{i}(x)\sigma_{1}\varphi_{i}(x) = g^{2}\sum_{i=1}^{N}
\varphi^{\dagger}_{i}(x)\sigma_{1}\chi_{i}(x)\approx\frac{\hat{m}}{\sqrt{2}}\end{equation}
yields for $m=0$, to leading order, the NJL-Lagrangian in which only quadratic fluctuations are kept
$${\cal L} = {\rm  i} \varphi ^{\dagger} \partial_{+} \varphi +{\rm  i} \chi ^{\dagger} \partial_{-}
\chi + \frac{{\rm i}}{\sqrt{2}}\left(  \varphi ^{\dagger} \tilde{\partial}_{\hat{m}}  \chi  +
\chi ^{\dagger} \partial_{\hat{m}} \varphi\right)\ .$$

Integrating out the constrained field $\chi$, the Hamiltonian of a system of non-interacting massive
fermions
\begin{equation}
\label{lcha}
H =\frac{{\rm i}}{2}\int {\rm d}^{3} x\,\varphi^{\dagger}\tilde{\partial}_{\hat{m}}\frac{1}
{\partial_{-}}\partial_{\hat{m}}\varphi
\end{equation}
is obtained. To determine the unknown mass parameter $\hat{m}$, we require the sum  in
Eq. (\ref{mass}) to be given by the limit of the vacuum expectation value of the sum over the
corresponding Heisenberg operators.
In the large $N$ limit, determination of the  spectrum and computation of  vacuum expectation
values of Heisenberg operators is simple. We obviously can use our above results with $m
\rightarrow \hat{m}$
and obtain, using Eq. (\ref{hei1}) and the asymptotics of the Bessel functions in the limit of
small spacelike $x^{2}$   the well known  gap equation of the NJL model
\begin{equation}
\label{hsc}
\hat{m}\left[\frac{g^2 N}{\pi^{2}} \left(\Lambda^{2} +\frac{\hat{m}^2}{4}\ln
\frac{\hat{m}^2}{\Lambda^2}\right)-1\right]=0\end{equation}
with  the cutoff $\Lambda$ defined by the point splitting procedure
\begin{displaymath}
\Lambda^2 = \frac{1}{-x^2}\ .
\end{displaymath}
This consistency condition is always solved trivially by $\hat{m}=0$. Beyond a critical coupling
(for fixed cutoff), Eq. (\ref{hsc}) has a solution with $\hat{m} \neq 0$ describing the phase with
spontaneously broken chiral symmetry. In ordinary coordinates, the solution with the lower energy
describes the stable phase. In light-cone quantization with its kinematically determined vacuum, the
vacuum energy cannot be determined variationally; stability can be checked either by evaluation of
the fluctuations (the NJL meson spectrum \cite{LOTY00}) or by calculation of the associated values of
 the effective potential (cf. \cite{MIRA93}). Since the effective potential is a Lorentz scalar, the
values obtained in ordinary coordinates are trivially reproduced for the solutions of the gap equation
(\ref{hsc}).\\
Identification of the chiral condensate with the limiting  VEV of light-cone Heisenberg operators is
crucial. Use of VEV's of Schr\"odinger operators (Eq. (\ref{sch1})) yields
\begin{equation}
  \label{njlv}
\frac{2N\hat{m}}{(2 \pi)^{3}} \int {\rm d}^{2}p_{\perp}
    \int^{\infty}_{0} \frac{{\rm d} p _{-}}{p_{-}}  =  \frac{\hat{m}}{g^2}
\end{equation}
which admits only the solution $\hat{m}=0$. \\
This  procedure also works in the 1+1 dimensional (chiral) GN model with its even more severe
infrared problems. Since in two dimensions $x^{+}=0$ denotes points on the light-``cone'', VEV's
of products of Schr\"odinger operators are necessarily singular;  again they are regularized by
point-splitting. The following substitution in Eq. (\ref{lc4})
\begin{displaymath}
  {\rm i}\partial_{m} \rightarrow -m \ , \quad x^{\perp} =0\ , \quad
 {\cal L}_{\rm int} = g^{2}(\varphi^{\dagger}\chi)(
  \chi^{\dagger}\varphi)
\end{displaymath}
defines the Gross-Neveu model in terms of the (one component) fields $\varphi,\chi$. The
relevant two-point function for non-interacting massive fermions is
$$ \langle 0|T(\varphi^{\dagger}(x)\chi(0))| 0 \rangle =\frac{{\rm i}m}{\sqrt{2}}
\int \frac{{\rm d}^{2}p}{(2 \pi)^{2}p_{-}} \; \frac{ {\rm e}^{{\rm i} p x}}{ p _{+}  -
\frac{m^{2} - {\rm i} \epsilon}{2 p_{-}}} = \frac{m}{\pi\sqrt{2}}K_{0}(m\sqrt{-x^{2}}) \ .     $$

The basic large $N$ limit now reads
$$g^{2}\sum_{i=1}^{N}\chi^{\dagger}_{i}(x)\varphi_{i}(x) = g^{2}\sum_{i=1}^{N}
\varphi^{\dagger}_{i}(x)\chi_{i}(x) )\approx -\frac{\hat{m}}{\sqrt{2}}$$
which yields, following the above arguments, the self-consistency equation \begin{equation}
  \label{sce}
 \hat{m}\left(1+\frac{Ng^{2}}{2\pi}\ln \frac{\hat{m}^{2}}{\Lambda^{2}}\right)
= 0\end{equation}
with
\begin{displaymath}
  \Lambda^{2} = \frac{4\, {\rm e}^{-2C}}{-x^{2}} \ .
\end{displaymath}
Eq. (\ref{sce}) again admits apart from $\hat{m}=0$ a non-trivial solution. This solution
defines the running of the coupling constant in terms of the physical mass $\hat{m}$; it  breaks
the 1+1 dimensional chiral symmetry
\begin{equation}
  \label{chirot2}
  \varphi(x)\rightarrow {\rm e}^{{\rm i}\alpha}\varphi(x)\ ,\quad \chi(x)\rightarrow
 {\rm e}^{-{\rm i}\alpha }\chi(x)\ . \\
\end{equation}
Once more, the solution is selected according to stability. In two dimensions the energy
density is a Lorentz scalar which, if regularized as $ -x^2 \rightarrow 0$ limit of Heisenberg
operators
\begin{eqnarray}
  \label{ende}
  \epsilon(\hat{m}) &=&  \langle 0|\left[-{\rm i}\chi^{\dagger}(x)\partial_{-}\chi(0)-g^{2}
(\varphi^{\dagger}(x)\chi(0))(
  \chi^{\dagger}(0)\varphi(x))\right]| 0 \rangle \nonumber\\& =& -\frac{\hat{m}^{2}}{4\pi}\ln
\frac{\hat{m}^{2}}{\Lambda^{2}}\left(1+\frac{Ng^{2}}{2\pi}\ln \frac{\hat{m}^{2}}
{\Lambda^{2}}\right) \ ,
\end{eqnarray}
agrees with the values of the effective potential at the stationary points, i.e., when the gap
equation is satisfied. In particular one obtains
\begin{displaymath}
\epsilon(\hat{m})- \epsilon(0) = -\frac{\hat{m}^{2}}{4\pi}\, .
\end{displaymath}
Thus for both the GN and the NJL model, light-cone quantization reproduces the well known
results of ordinary quantization. Within these models, the simplicity of the light-cone description
is not spoiled by a dynamical symmetry breakdown.  \\
Our resolution of the triviality problem of the light-cone vacuum differs from the outset from
previous attempts  which have focused on the VEV's of Schr\"odinger operators. Regularization
of VEV's leading to expressions
like in Eq. (\ref{njlv}) offers the possibility for introducing dynamical dependences
into these purely kinematical objects. In the context of the NJL model, rules for regularization
have been proposed by which the value of the chiral condensate obtained in ordinary quantization
could be reproduced \cite{Dietmaier89,Bentz99, Itakura00}. However it is difficult to see how,
by such rules, the difference in the dynamics of broken and unbroken phase could be accounted
for or how covariance in the evaluation of the corresponding correlation functions for non-vanishing
spacelike separations could be respected (cf. \cite{ Tsujimaru98}). In the approach we have
described, non trivial vacuum properties are associated with products of Heisenberg operators
in the equal light-cone time limit. Unlike in standard quantization schemes, VEV's determined
in such a limiting procedure do not agree with VEV's of products of the corresponding
Schr\"odinger operators and it is only the latter ones whose VEV's are trivial. It is by this
subtle distinction between Schr\"odinger operators and the equal time limit of Heisenberg
operators that condensates serving as order parameters for spontaneously broken symmetries
can be defined despite the triviality of the ground state. From this point of view, the successful
evaluation of the chiral condensate of the 't~Hooft model in \cite{Zhitnitsky} becomes plausible;
it avoids completely light-cone Schr\"odinger operators and uses general properties of VEV's
which on the light-cone can be attributed only to expectation values of limits of Heisenberg operators.
In a similar vein one can understand why the condensate issue could be bypassed in a
light-cone calculation of fermion-antifermion scattering and bound states in
the GN model \cite{Thies93}. Finally, in the correct determination of the chiral condensate
of the Schwinger model in \cite{Nakawaki00}, the use of Heisenberg operators and
point splitting was an essential element.
\\
With the identification of VEV's of appropriate limits of Heisenberg operators as the relevant
quantities for definition of order parameters, the standard tools of analyzing the effects of broken
symmetries become available to light-cone quantization \cite{LOTY00}. Ward identities can be
derived  and their consequences such as the  Gell-Mann, Oakes, Renner relation  \cite{GOR} can
be studied within light-cone quantization; perturbative treatments of explicit symmetry violations
become amenable to the light-cone approach. Although our analysis has focused on fermionic
theories, the extension to bosons is straightforward unless the VEV to be considered is linear
in the field operator. In this particular case, as has been advocated in various studies
(cf. \cite{Maskawa76,Harindranath87, Heinzl92,Bender93,Pinsky94}) the dynamics of a
single (zero) mode may require a special treatment. \\
For light-cone studies of QCD the distinction
between VEV's of Schr\"odinger operators and of limits of Heisenberg operators will be significant
not only for the description of the quark condensate but also for the gluon condensate which is
quadratic and of higher order in the gauge fields. Extension to gauge theories introduces a novel
dynamical element into the discussion.  Definition of non-trivial vacuum expectation values in
light-cone quantization requires splitting in light-cone time; in turn, gauge invariance requires, in
light-cone gauge $A_{-}=0$, associated gauge strings to be introduced whose effects are
expected to be enhanced by the infrared ($p_{-}=0$) singularity characteristic for light-cone
quantization.

\end{document}